\documentclass[titlepage,12pt]{article}
\usepackage[dvips]{graphicx}

\begin{document}

\title{Effects of a mixed vector-scalar screened Coulomb potential for spinless
particles}
\date{}
\author{Antonio S. de Castro \\
\\
UNESP - Campus de Guaratinguet\'{a}\\
Departamento de F\'{\i}sica e Qu\'{\i}mica\\
Caixa Postal 205\\
12516-410 Guaratinguet\'{a} SP - Brasil\\
\\
E-mail address: castro@feg.unesp.br (A.S. de Castro)}
\date{}
\maketitle

\begin{abstract}
The problem of a spinless particle subject to a general mixing of vector and
scalar screened Coulomb potentials in a two-dimensional world is analyzed
and its bounded solutions are found. Some unusual results, including the
existence of a bona fide solitary zero-eigenmode solution, are revealed for
the Klein-Gordon equation. The cases of pure vector and scalar potentials,
already analyzed in previous works, are obtained as particular cases.
\end{abstract}

\section{Introduction}

In a two-dimensional space-time the screened Coulomb potential ($\sim
e^{-|x|/\lambda }$) \thinspace has been analyzed and its analytical
solutions have been found for the Dirac equation with vector \cite{ada},
scalar \cite{kag3} and pseudoscalar \cite{asc} couplings and for the
Klein-Gordon (KG) equation with vector \cite{kag1} and scalar \cite{kag2}
couplings. As has been emphasized in Refs. \cite{kag1} and \cite{kag2}, the
solution of the KG equation with this sort of potential may find
applications in the study of pionic atoms, doped Mott insulators, doped
semiconductors, interaction between ions, quantum dots surrounded by a
dielectric or a conducting medium, protein structures, etc.

In the present work the problem of a spinless particle in the background of
a screened Coulomb potential is considered with a general mixing of vector
and scalar Lorentz structures. This sort of mixing beyond its potential
physical applications, shows to be a powerful tool to obtain a deeper
insight about the nature of the KG equation and its solutions. The problem
is mapped into an exactly solvable Sturm-Liouville problem of a Schr\"{o}%
dinger-like equation with an effective symmetric Morse-like potential, or an
effective screened Coulomb potential in particular circumstances. The cases
of pure vector and scalar potentials, already analyzed in \cite{kag1}-\cite
{kag2}, are obtained as particular cases.

In the presence of vector and scalar potentials the 1+1 dimensional
time-independent KG equation for a spinless particle of rest mass $m$ reads

\begin{equation}
-\hbar ^{2}c^{2}\,\frac{d^{2}\psi }{dx^{2}}+\left( mc^{2}+V_{s}\right)
^{2}\psi =\left( E-V_{v}\right) ^{2}\psi  \label{1b}
\end{equation}

\noindent where $E$ is the energy of the particle, $c$ is the velocity of
light and $\hbar $ is the Planck constant. The vector and scalar potentials
are given by $V_{v}$ and $V_{s}$, respectively. The subscripts for the terms
of potential denote their properties under a Lorentz transformation: $v$ for
the time component of the 2-vector potential and $s$ for the scalar term. It
is worth to note that the KG equation is covariant under $x\rightarrow -x$
if $V_{v}(x)$ and $V_{s}(x)$ remain the same. Also note that $\psi $ remains
invariant under the simultaneous transformations $E\rightarrow -E$ and $%
V_{v}\rightarrow -V_{v}$. Furthermore, for $V_{v}=0$, the case of a pure
scalar potential, the negative- and positive-energy levels are disposed
symmetrically about $E=0$.

The KG equation can also be written as
\begin{equation}
H_{eff}\psi =-\frac{\hbar ^{2}}{2m}\,\psi ^{\prime \prime }+V_{eff}\,\psi
=E_{eff}\,\psi  \label{1c}
\end{equation}

\noindent where
\begin{equation}
E_{eff}=\frac{E^{2}-m^{2}c^{4}}{2mc^{2}},\quad V_{eff}=\frac{%
V_{s}^{2}-V_{v}^{2}}{2mc^{2}}+V_{s}+\frac{E}{mc^{2}}\,V_{v}  \label{1d}
\end{equation}

\noindent From this one can see that for potentials which tend to $\pm
\infty $ as $|x|\rightarrow \infty $ it follows that $V_{eff}\rightarrow
\left( V_{s}^{2}-V_{v}^{2}\right) /\left( 2mc^{2}\right) $, so that the KG
equation furnishes a purely discrete (continuum) spectrum for $%
|V_{s}|>|V_{v}|$ ($|V_{s}|<|V_{v}|$). On the other hand, if the potentials
vanish as $|x|\rightarrow \infty $ the continuum spectrum is omnipresent but
the necessary conditions for the existence of a discrete spectrum is not an
easy task for general functional forms. The boundary conditions on the
eigenfunctions come into existence by demanding that the effective
Hamiltonian given (\ref{1c}) is Hermitian, viz.

\begin{equation}
\int_{a}^{b}dx\;\psi _{n}^{*}\left( H_{eff}\psi _{n^{^{\prime }}}\right)
=\int_{a}^{b}dx\;\left( H_{eff}\psi _{n}\right) ^{*}\psi _{n^{^{\prime }}}
\label{22-1}
\end{equation}

\noindent where $\psi _{n}$ is an eigenfunction corresponding to an
effective eigenvalue $\left( E_{eff}\right) _{n}$ and $\left( a,b\right) $
is the interval under consideration. In passing, note that a necessary
consequence of Eq. (\ref{22-1}) is that the eigenfunctions corresponding to
distinct effective eigenvalues are orthogonal. It can be shown that (\ref
{22-1}) is equivalent to

\begin{equation}
\left[ \psi _{n}^{*}\frac{d\psi _{n^{^{\prime }}}}{dx}-\frac{d\psi _{n}^{*}}{%
dx}\psi _{n^{^{\prime }}}\right] _{x=a}^{x=b}=0  \label{22-2}
\end{equation}
\noindent In the nonrelativistic approximation (potential energies small
compared to $mc^{2}$ and $E\simeq mc^{2}$) Eq. (\ref{1b}) becomes

\begin{equation}
\left( -\frac{\hbar ^{2}}{2m}\frac{d^{2}}{dx^{2}}+V_{v}+V_{s}\right) \psi
=\left( E-mc^{2}\right) \psi  \label{1e}
\end{equation}

\noindent so that $\psi $ obeys the Schr\"{o}dinger equation with binding
energy equal to $E-mc^{2}$ without distinguishing the contributions of
vector and scalar potentials.

It is remarkable that the KG equation with a scalar potential, or a vector
potential contaminated with some scalar coupling, is not invariant under $%
V\rightarrow V+const.$, this is so because only the vector potential couples
to the positive-energies in the same way it couples to the negative-ones,
whereas the scalar potential couples to the mass of the particle. Therefore,
if there is any scalar coupling the absolute values of the energy will have
physical significance and the freedom to choose a zero-energy will be lost.
It is well known that a confining potential in the nonrelativistic approach
is not confining in the relativistic approach when it is considered as a
Lorentz vector. It is surprising that relativistic confining potentials may
result in nonconfinement in the nonrelativistic approach. This last
phenomenon is a consequence of the fact that vector and scalar potentials
couple differently in the KG equation whereas there is no such distinction
among them in the Schr\"{o}dinger equation. This observation permit us to
conclude that even a ``repulsive'' potential can be a confining potential.
The case $V_{v}=-V_{s}$ presents bounded solutions in the relativistic
approach, although it reduces to the free-particle problem in the
nonrelativistic limit. The attractive vector potential for a particle is, of
course, repulsive for its corresponding antiparticle, and vice versa.
However, the attractive (repulsive) scalar potential for particles is also
attractive (repulsive) for antiparticles. For $V_{v}=V_{s}$ and an
attractive vector potential for particles, the scalar potential is
counterbalanced by the vector potential for antiparticles as long as the
scalar potential is attractive and the vector potential is repulsive. As a
consequence there is no bounded solution for antiparticles. For $V_{v}=0$
and a pure scalar attractive potential, one finds energy levels for
particles and antiparticles arranged symmetrically about $E=0$. For $%
V_{v}=-V_{s}$ and a repulsive vector potential for particles, the scalar and
the vector potentials are attractive for antiparticles but their effects are
counterbalanced for particles. Thus, recurring to this simple standpoint one
can anticipate in the mind that there is no bound-state solution for
particles in this last case of mixing.

\section{The mixed vector-scalar screened Coulomb potential}

Now let us focus our attention on scalar and vector potentials in the form
\begin{equation}
V_{s}=-\frac{g_{s}}{2\lambda }\exp \left( -\frac{|x|}{\lambda }\right)
,\quad V_{v}=-\frac{g_{v}}{2\lambda }\exp \left( -\frac{|x|}{\lambda }\right)
\label{12}
\end{equation}
\noindent where the coupling constants, $g_{s}$ and $g_{v}$, are
dimensionless real parameters and $\lambda $, related to the range of the
interaction, is a positive parameter. In this case the second equation of (%
\ref{1d}) transmutes into
\begin{equation}
V_{eff}=V_{1}\exp \left( -\frac{|x|}{\lambda }\right) +V_{2}\exp \left( -2%
\frac{|x|}{\lambda }\right)  \label{12b}
\end{equation}
where
\begin{equation}
V_{1}=-\frac{1}{2\lambda }\left( g_{s}+\frac{E}{mc^{2}}\,g_{v}\right) ,\quad
V_{2}=\frac{g_{s}^{2}-g_{v}^{2}}{8\lambda ^{2}mc^{2}}  \label{14}
\end{equation}

\noindent Therefore, one has to search for bounded solutions in an effective
symmetric Morse-like potential for $g_{s}^{2}\neq g_{v}^{2}$, or screened
Coulomb potential for $g_{s}^{2}=g_{v}^{2}$. The KG eigenvalues are obtained
by inserting the effective eigenvalues into the first equation of (\ref{1d}%
). Since the effective potential is even under $x\rightarrow -x$, the KG
eigenfunction can be expressed as a function of definite parity. Thus, we
can concentrate our attention on the positive half-line and impose boundary
conditions on $\psi $ at $x=0$ and $x=+\infty $. From (\ref{22-2}) one can
see that in addition to $\psi \left( \infty \right) =0$, the boundary
conditions can be met in two distinct ways: the odd function obeys the
Neumann condition at the origin ($d\psi /dx|_{x=0}=0$) whereas the even
function obeys the Dirichlet condition ($\psi \left( 0\right) =0$) .

Note carefully that the potentials $V_{s}$ and $V_{v}$ vanish as $%
|x|\rightarrow \infty $ and the KG equation can furnish a discrete spectrum
when $V_{1}<0$ and $V_{2}\geq 0$, or $V_{1}<|V_{2}|$ and $V_{2}<0$. Only in
those circumstances the effective potentials present potential-well
structures permitting bounded solutions in the range $|E|<mc^{2}$. The
eigenenergies in the range $|E|>mc^{2}$ correspond to the continuum.

Now we move to consider a quantitative treatment of our problem by
considering the two distinct classes of effective potentials.

\subsection{\noindent The effective screened Coulomb potential ($%
g_{s}^{2}=g_{v}^{2}$)}

For this class of effective potential, the discrete spectrum arises when $%
V_{1}<0$ and $V_{2}=0$, corresponding to $g_{s}\left[ 1+\mathrm{sgn}\left(
g_{v}\right) E/\left( mc^{2}\right) \right] >0$ and $g_{s}=|g_{v}|$.
Defining the dimensionless quantities
\begin{eqnarray}
y &=&y_{0}\exp \left( -\frac{|x|}{2\lambda }\right) ,\qquad y_{0}=\frac{2}{%
\hbar }\sqrt{\lambda mg_{s}\left[ 1+\frac{E}{mc^{2}}\,\mathrm{sgn}\left(
g_{v}\right) \right] }  \nonumber \\
&&  \label{20} \\
\mu  &=&\frac{2\lambda mc}{\hbar }\sqrt{1-\frac{E^{2}}{m^{2}c^{4}}}
\nonumber
\end{eqnarray}

\noindent and using (\ref{1c})-(\ref{1d}) and (\ref{12b})-(\ref{14}) one
obtains the differential Bessel equation
\begin{equation}
y^{2}\psi ^{\prime \prime }+y\psi ^{\prime }+\left( y^{2}-\mu ^{2}\right)
\psi =0  \label{21}
\end{equation}

\noindent where the prime denotes differentiation with respect to $y$. The
solution finite at $y=0$ ($|x|=\infty $) is given by the Bessel function of
the first kind and order $\mu $ \cite{abr}:
\begin{equation}
\psi (y)=N_{\mu }\,J_{\mu }(y)  \label{21f}
\end{equation}
\noindent where $N_{\mu }$ is a normalization constant. In fact, the
normalizability of $\psi $ demands that the integral $%
\int_{0}^{y_{0}}y^{-1}|J_{\mu }(y)|^{2}dy$ must be convergent. Since $J_{\mu
}(y)$ behaves as $y^{\mu }$ at the lower limit, one can see that $\mu \geq
1/2$ so that square-integrable KG eigenfuntions are allowed only if $\lambda
\geq \lambda _{c}/4$, where $\lambda _{c}=\hbar /(mc)$ is the Compton
wavelength. The boundary conditions at $x=0$ ($y=y_{0}$) imply that
\begin{equation}
\begin{array}{ll}
\frac{dJ_{\mu }(y)}{dy}|_{y=y_{0}}=0, & {\textrm{for even states}} \\
&  \\
J_{\mu }(y_{0})=0, & \textrm{for odd states}
\end{array}
\label{22}
\end{equation}

\noindent Since the KG eigenenergies are dependent on $\mu $ and $y_{0}$, it
follows that Eq. (\ref{22}) is a quantization condition. The allowed values
for the parameters $\mu $ and $y_{0}$, and $E$ as an immediate consequence,
are determined by solving Eq. (\ref{22}). The oscillatory character of the
Bessel function and the finite range for $y$ ($0<y\leq y_{0}$) imply that
there is a finite number of discrete KG eigenenergies. Of course, the number
of bound-state solutions increases as $y_{0}$ is increased. As we let $%
\lambda \rightarrow \lambda _{c}/4$, it can now be realized that the energy
levels approach $E=0$ and tend to disappear one after another. It happens
that an isolated zero-energy solution survives when $\lambda =\lambda _{c}/4
$, regardless the relative values of the coupling constants (recall that $%
g_{s}>0$).

The roots of $J_{\mu }(y)$ and $J_{\mu }^{\prime }(y)$ are listed in tables
of Bessel functions only for a few special values of $\mu $. A bit of time
and effort can be saved in the numerical calculation of the roots of $J_{\mu
}^{\prime }(y)$ if one uses the recurrence relation $J_{\mu -1}-J_{\mu
+1}=2J_{\mu }^{\prime }$, in such a manner that the quantization condition
for even states translates into $J_{\mu +1}(y_{0})=J_{\mu -1}(y_{0})$.

When $g_{s}=g_{v}$ ($g_{s}=-g_{v}$) the single-well potential is deeper
(shallower) for positive-energy levels than that one for negative-energy
levels. Thus, the capacity to hold bound states depends on the sign of the
eigenenergy and one might expect that the number of positive (negative)
energy levels is greater than the number of negative (positive) energy
levels. By the way, the positive (negative) energy solutions are not to be
promptly identified with the solutions for particles (antiparticles).
Rather, whether it is positive or negative, an eigenenergy can be
unambiguously identified with a bounded solution for a particle
(antiparticle) only by observing if the energy level emerges from the upper
(lower) continuum.

\begin{figure}[th]
\begin{center}
\includegraphics[width=9cm, angle=270]{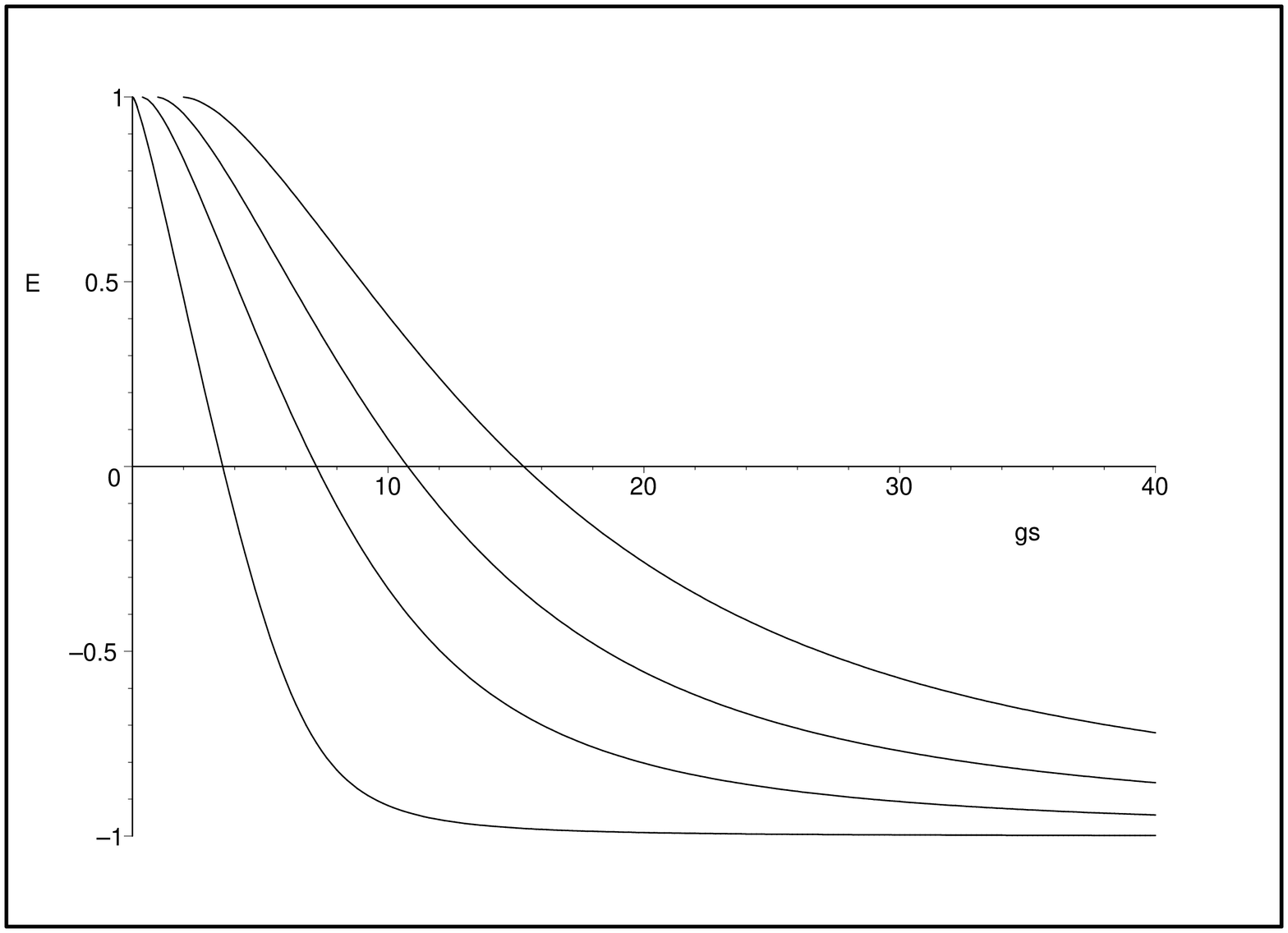}
\end{center}
\par
\vspace*{-0.1cm}
\caption{KG eigenenergies for the four lowest energy levels as a function of
$g_{s}$ for $g_{v}=g_{s}$ ($\lambda =2\lambda_{c}$ and $m=c=\hbar=1$). }
\label{Fig1}
\end{figure}

The KG eigenenergies are plotted in Fig. \ref{Fig1} for the four lowest
bound states as a function of $g_{s}$ for $g_{v}=g_{s}$ and $\lambda
=2\lambda _{c}$. The eigenenergies for $g_{v}=-g_{s}$ can be obtained by
changing $E$ by $-E$, as mentioned before. Note that in the case illustrated
in Fig. \ref{Fig1} the eigenenergies correspond to bounded solutions for
particles. There are no energy levels for antiparticles. Also, note that the
energy level corresponding to the ground-state solution ($\psi $ even)
always makes its appearance and that the number of energy levels grows with $%
g_{s}$. The spectrum consists of a finite set of energy levels of alternate
parities. The nonrelativistic limit is only viable for $g_{s}=g_{v}$ whereas
the case $g_{s}=-g_{v}$ is an essentially relativistic problem. Furthermore,
one has $E\simeq mc^{2}$ as long as $g_{s}\ll 1$.

\subsection{\noindent The effective Morse-like potential ($g_{s}^{2}\neq
g_{v}^{2}$)}

For this class, the existence of bound-state solutions permits us to
distinguish two subclasses: 1) $V_{1}<0$ and $V_{2}>0$, corresponding to $%
g_{s}+g_{v}E/\left( mc^{2}\right) >0$ and $g_{s}>|g_{v}|$; 2) $V_{1}<|V_{2}|$
and $V_{2}<0$, corresponding to $g_{s}+g_{v}E/\left( mc^{2}\right) >-\left(
g_{v}^{2}-g_{s}^{2}\right) /\left( 4mc^{2}\right) $ and $|g_{s}|<|g_{v}|$.
Included into the first (second) subclass is the case of a pure scalar
(vector) coupling. The first subclass, as well as the second one on the
condition that $g_{v}>|g_{s}|$, contain the nonrelativistic theory as a
limiting case. On the contrary, i.e. $g_{s}<|g_{v}|$ and $g_{v}<0$, the
theory is essentially relativistic. Let us define
\begin{eqnarray}
z &=&z_{0}\exp \left( -\frac{|x|}{\lambda }\right) ,\qquad z_{0}=\frac{\sqrt{%
g_{s}^{2}-g_{v}^{2}}}{\hbar c}  \nonumber \\
&&  \label{24} \\
\rho &=&\frac{\lambda m}{\hbar ^{2}z_{0}}\left( g_{s}+\frac{E}{mc^{2}}%
\,g_{v}\right) ,\qquad \nu =\frac{\lambda mc}{\hbar }\sqrt{1-\frac{E^{2}}{%
m^{2}c^{4}}}  \nonumber
\end{eqnarray}

\noindent so that
\begin{equation}
z\psi ^{\prime \prime }+\psi ^{\prime }+\left( -\frac{z}{4}-\frac{\nu ^{2}}{z%
}+\rho \right) \psi =0  \label{16}
\end{equation}

\noindent Note that $\psi $ is a function of complex variable, $z$, if $%
g_{s}^{2}<g_{v}^{2}$. Following the steps of Refs. \cite{kag1} and \cite
{kag2}, we make the transformation $\psi =z^{-1/2}\phi $ to obtain the
Whittaker equation \cite{abr}:
\begin{equation}
\phi ^{\prime \prime }+\left( -\frac{1}{4}+\frac{\rho }{z}+\frac{1/4-\nu ^{2}%
}{z^{2}}\right) \phi =0  \label{16a}
\end{equation}
whose solution vanishing at the infinity is written as $\phi =N\,z^{\nu
+1/2}e^{-z/2}M(a,b,z)$, where $N$ is a normalization constant and $M$ is the
regular confluent hypergeometric function \noindent with

\begin{equation}
a=\nu +\frac{1}{2}-\rho ,\qquad b=2\nu +1  \label{17}
\end{equation}

\noindent Thus,
\begin{equation}
\psi =Nz^{\nu }e^{-z/2}M(a,b,z)  \label{18}
\end{equation}

\noindent For this class of effective potential there is no restriction on
the size of $\lambda $ in order to make the existence of a bounded solution
possible as there is for the previous class. The KG eigenfunction is
normalizable for any $\nu $ as easy inspection shows. Therefore, one can
think of a very short-ranged potential in the sense of $\lambda \rightarrow
0 $, i.e. a potential approaching the $\delta $-function. Indeed, this sort
of limit has already been realized by Dom\'{i}nguez-Adame and Rodr\'{i}guez
\cite{ada} for the case of a pure vector potential. From Eq. (\ref{18}) one
can see now that the boundary conditions at $x=0$ ($z=z_{0}$) imply into the
quantization conditions
\begin{equation}
\begin{array}{ll}
\frac{M\left( a+1,b+1,z_{0}\right) }{M\left( a,b,z_{0}\right) }=\frac{%
z_{0}-2\nu }{2\frac{a}{b}z_{0}}, & {\textrm{for even states}} \\
&  \\
M(a,b,z_{0})=0, & {\textrm{for odd states}}
\end{array}
\label{19}
\end{equation}

If $g_{s}$ happens to vanish, the spectrum will only consist of positive
(negative) energy levels for $g_{v}>0$ ($g_{v}<0$). If $g_{s}\neq 0$,
though, the spectrum may acquiesce both signs of eigenenergies. The presence
of both signs of eigenenergies depends, of course, on the relative strength
between the vector and scalar potentials. When $g_{v}=0$ the negative- and
positive-energy levels are disposed symmetrically about $E=0$, as commented
before, so that there are as many positive-energy levels as negative ones.
In the case $g_{s}>|g_{v}|$ it is reasonable to expect a two-fold degeneracy
as $g_{s}/|g_{v}|\rightarrow \infty $ due to the double-well structure with
an infinitely high barrier potential between the wells. That degeneracy in
an one-dimensional quantum-mechanical problem is due to the fact that even
eigenfunctions tend to vanish at the origin as $g_{s}/|g_{v}|\rightarrow
\infty $.

The Fig. \ref{Fig2} illustrates the four lowest states of the spectrum for
this class of effective potential as a function of $g_{s}/|g_{v}|$ with $%
g_{v}>0$. The energy level corresponding to the ground-state solution ($\psi
$ even) always makes its appearance. As before, the eigenenergies for $%
g_{v}<0$ can be obtained by replacing $E$ by $-E$ (recall that $%
g_{s}>-|g_{v}|$). Note that only bounded solutions for particles are present
for $g_{s}/|g_{v}|<1$ (even if $E<0$) and that a new branch of solutions
corresponding to antiparticles emerges from the lower continuum as $%
g_{s}/|g_{v}|$ increases starting from $g_{s}/|g_{v}|=1$. In that last case
as $g_{s}/|g_{v}|\rightarrow \infty $ the even and odd parities solutions
tend to be degenerate and the spectrum tends to exhibit a symmetry about $%
E=0 $.

\section{Conclusions}

Using the same method used in prior works, we have succeed in the proposal
of searching the solution for a more general screened Coulomb potential with
the KG equation. An opportunity was given by that generalization to analyze
some aspects of the KG equation which would not be feasibly only with the
special cases already approached in the literature. Thus, the use of the
mixing of vector and scalar Lorentz structures for other kinds of potentials
may lead to a better understanding of the KG equation and its solutions.
Free from doubt, this sort of mixing also deserves to be explored with the
Dirac equation.

\begin{figure}[th]
\begin{center}
\includegraphics[width=9cm, angle=270]{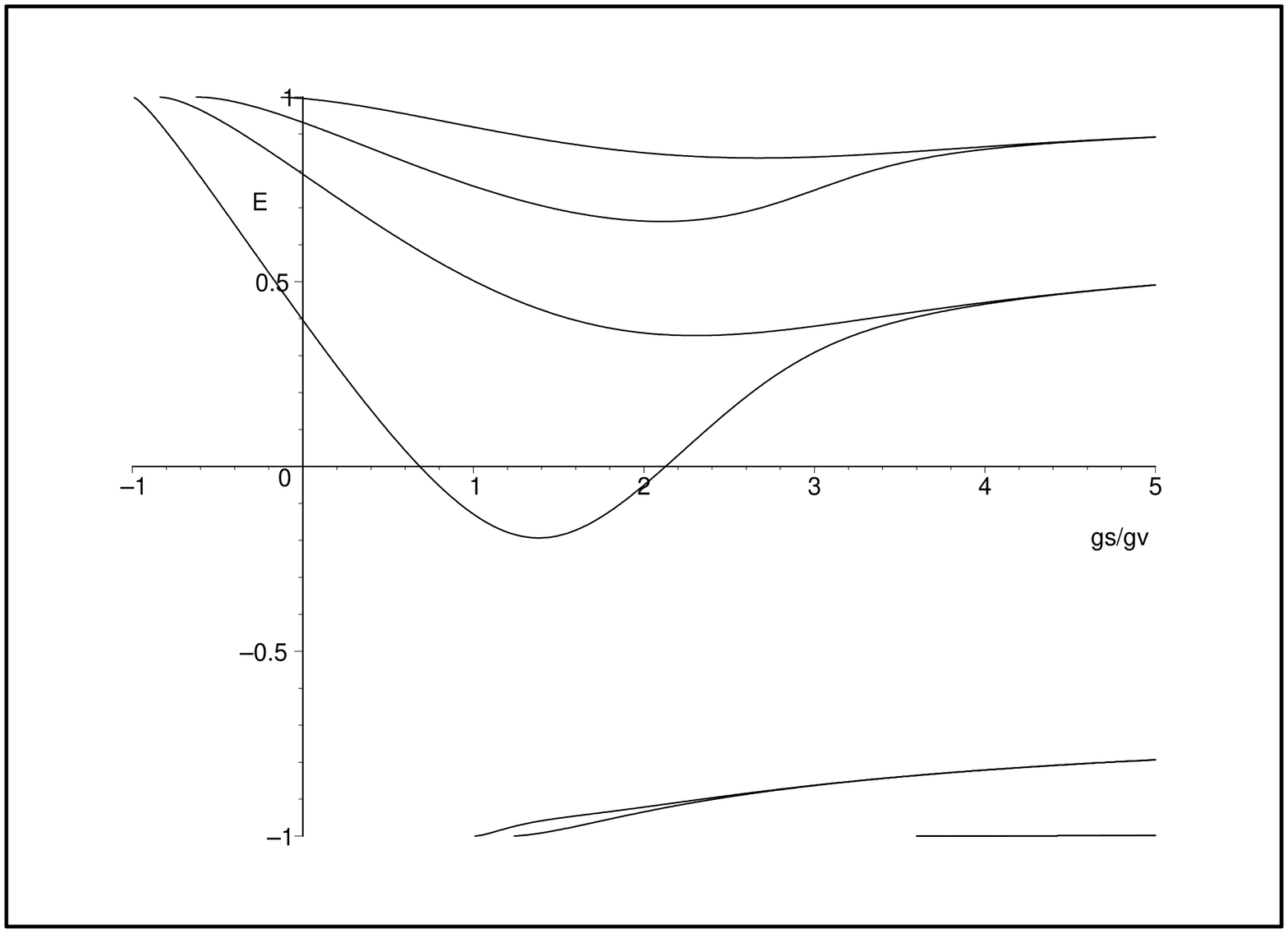}
\end{center}
\par
\vspace*{-0.1cm}
\caption{KG eigenenergies for the four lowest energy levels as a function of
$g_{s}/g_{v}$ ($g_{v}=4*$, $\lambda =2\lambda_{c}$ and $m=c=\hbar=1$). }
\label{Fig2}
\end{figure}

It is worthwhile to mention that the solutions of the KG equation with a
screened Coulomb potential present a continuous transition as the ratio $%
g_{s}/g_{v}$ varies. However, a phase transition occurs when $|g_{s}/g_{v}|=1
$. Although the phase transition does not always show its face for the KG
eigenenergies (observe carefully the continuity of the KG eigenenergies for
the particle energy levels in Fig. \ref{Fig2}), it clearly shows it for the
KG eigenfunctions (note, for instance, that the behaviour of the KG
eigenfunction for both classes of effective potentials differ at the
neighborhood of the origin).

Finally, we draw attention to the fact that no matter how strong the
potentials are, as far as $g_{s}\geq -|g_{v}|$, the energy levels for
particles (antiparticles) never dive into the lower (upper) continuum. Thus
there is no room for the production of particle-antiparticle pairs. This all
means that Klein\'{}s paradox never comes to the scenario.

\vspace{1in}

\noindent{\textbf{Acknowledgments} }

This work was supported in part by means of funds provided by CNPq and
FAPESP.

\newpage

\newpage

\begin{thebibliography}{9}
\bibitem{ada}  F. Dom\'{i}nguez-Adame, A. Rodr\'{i}guez, Phys. Lett. A 198
(1995) 275.

\bibitem{kag3}  B.A. Kagali, N.A. Rao, V. Sivramkrishna, Mod. Phys. Lett. A
17 (2002) 2049.

\bibitem{asc}  A.S. de Castro, Ann. Phys. (N.Y.), in press.

\bibitem{kag1}  N.A. Rao, B.A. Kagali, Phys. Lett. A 296 (2002) 192.

\bibitem{kag2}  N.A. Rao, B.A. Kagali, V. Sivramkrishna, Int. J. Mod. Phys.
A 17 (2002) 4793.

\bibitem{abr}  M. Abramowitz, I.A. Stegun, Handbook of Mathematical
Functions, Dover, Toronto, 1965.
\end{thebibliography}
\end{document}